\documentclass[fleqn,usenatbib]{mnras}
\usepackage{newtxtext,newtxmath}

\usepackage[T1]{fontenc}
\usepackage{ae,aecompl}
\usepackage{verbatim}

\usepackage{graphicx}	
\usepackage{xspace}        

\newcommand{\target}{J203248.825$+$404804.18\xspace}

\newcommand{\Ni}{\textit{NICER}\xspace}
\newcommand{\flux}{\,erg\,s$^{-1}$\,cm$^{-2}$\xspace} 

\title[Innermost jet in the ULX Cyg X-3]{The innermost jet in the hidden ultra-luminous X-ray source Cygnus X-3}
\author[J. Yang et al.]{Jun Yang$^{1}$\thanks{E-mail: jun.yang@chalmers.se}, Federico Garc\'ia$^{2}$\thanks{E-mail: fgarcia@iar.unlp.edu.ar},
Santiago del Palacio$^{1}$,
Ralph Spencer$^{3}$,
Zsolt Paragi$^{4}$,
Noel Castro~Segura$^{5}$,
\and
Biping Gong$^{6}$,
Hongmin Cao$^{7}$
and Wen Chen$^{8, 9}$
\\
$^{1}$Department of Space, Earth and Environment, Chalmers University of Technology, Onsala Space Observatory, SE-439 92 Onsala, Sweden \\
$^{2}$Instituto Argentino de Radioastronom\'ia (CCT La Plata, CONICET; CICPBA; UNLP), C.C.5, (1894) Villa Elisa, Buenos Aires, Argentina \\
$^{3}$Jodrell Bank Centre for Astrophysics, Department of Physics and Astronomy, The University of Manchester, Oxford Rd., Manchester M13 9PL, UK \\
$^{4}$Joint Institute for VLBI ERIC (JIVE), Oude Hoogeveensedijk 4, 7991 PD Dwingeloo, The Netherlands \\
$^{5}$School of Physics and Astronomy, University of Southampton, Southampton
SO17 1BJ, UK \\
$^{6}$School of Physics, Huazhong University of Science and Technology, Wuhan 430074, China \\
$^{7}$School of Electronic and Electrical Engineering, Shangqiu Normal University, 298 Wenhua Road, Shangqiu, Henan 476000, China \\
$^{8}$Yunnan Observatories, Chinese Academy of Sciences, Kunming 650216, Yunnan, China\\
$^{9}$Key Laboratory for the Structure and Evolution of Celestial Objects, Chinese Academy of Sciences, Kunming 650216, China\\
}
\date{Accepted 2023 XXX. Received 2023 YYY; in original form 2023 ZZZ}

\pubyear{2023}

\begin{document}
\label{firstpage}
\pagerange{\pageref{firstpage}--\pageref{lastpage}}
\maketitle
\begin{abstract}
Cygnus~X-3 is a high-mass X-ray binary with a compact object accreting matter from a Wolf-Rayet donor star. Recently, it has been revealed by the \textit{Imaging X-ray Polarimetry Explorer} (\textit{IXPE}) as a hidden Galactic ultra-luminous X-ray (ULX) source with a luminosity above the Eddington limit along the direction of a narrow (opening angle $\la32\degr$) funnel. In between the \textit{IXPE} observations, we observed Cyg~X-3 with the European VLBI (very long baseline interferometry) Network at 22~GHz and the \textit{NICER} X-ray instrument. To probe possible relations between the X-ray funnel and the potential radio jet from the ULX, we analyzed the simultaneous multi-wavelength data. Our high-resolution VLBI image reveals an elongated structure with a position angle of $-3\fdg2\pm0\fdg4$, accurately perpendicular to the direction of the linear X-ray polarization. Because Cyg~X-3 was in the radio quiescent state on 2022 November 10, we identify the mas-scale structure as the innermost radio jet. The finding indicates that the radio jet propagates along and within the funnel. Moreover, the jet is marginally resolved in the transverse direction. This possibly results from the strong stellar winds and the rapid orbital motion of the binary system.     
\end{abstract}

\begin{keywords}
stars: individual: Cygnus~X-3 -- stars: jets -- X-rays: binaries -- radio continuum: stars
\end{keywords}




\section{Introduction}
\label{sec1}

\begin{figure}
\centering
\includegraphics[width=\columnwidth]{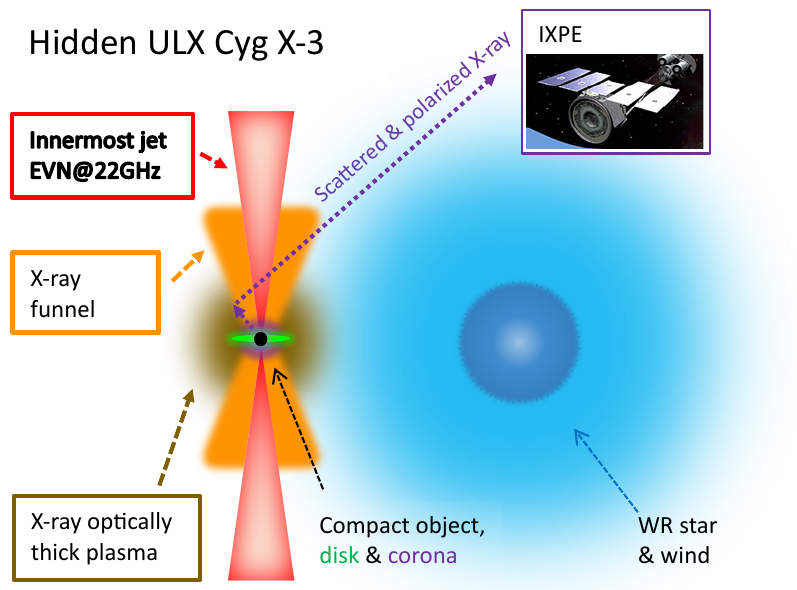}
\caption{
Schematic view of the inner structure of the Galactic hidden ULX Cyg X-3 in the radio quiescent state. }
\label{fig:cartoon}
\end{figure}

Cygnus~X-3 (Cyg~X-3) is a high-mass X-ray binary system with a compact object accreting matter from a Wolf-Rayet donor star \citep{vanKerkwijk1992}. The binary system has an orbital period of 4.8~h \citep{Parsignault1972}. Cyg~X-3 was discovered more than five decades ago \citep{Giacconi1967}. To date, it is an open question whether the compact object is a black hole or a neutron star. Recently, Cyg~X-3 has been recognized by \citet{Veledina2023} as a hidden Galactic ultra-luminous X-ray (ULX) source \citep[the first Galactic ULX was SS433, e.g.][]{Fabrika2015} with the \textit{IXPE} observations. This makes it a precious target to unveil the physics of the more distant extragalactic ULX sources \citep[cf. reviews by][]{Kaaret2017, King2023}. The X-ray emission of Cyg~X-3 shows a very high ($\sim$25\%) and nearly energy-independent linear polarization and a relatively low apparent luminosity because of strong Thomson scattering from a surrounding optically thick funnel. If the X-ray funnel is indeed quite narrow \citep[opening angle $\la32\degr$,][]{Veledina2023}, its origin would be tightly related to potential jet activity. The proposed inner structure of Cyg X-3 is also shown in Fig.~\ref{fig:cartoon}.

Cyg~X-3 is also a highly interesting microquasar. Microquasars are a subclass of Galactic X-ray binaries that launch relativistic jets \citep{Mirabel1994}. Thanks to their much smaller sizes, these scaled-down versions of extragalactic quasars allow us to study accretion and ejection activity at various accretion rates on short timescales \citep[e.g.][]{Fender2009}. As a microquasar, Cyg~X-3 has a persistent radio counterpart with typical flux densities $\sim$100~mJy at 1--10 GHz \citep[e.g.][]{Trushkin2017} and 20--100~mJy at 22~GHz \citep[e.g.][]{Tsuboi2008} in the quiescent radio state \citep[e.g][]{Szostek2008}. Frequently, it undergoes bright outbursts with peak flux densities reaching $\sim$20~Jy \citep[e.g.][]{Gregory1972, Waltman1994, Tudose2010, Kim2013, Egron2017, Kim2020, Broderick2021, Spencer2022}. During some giant outbursts, high-resolution observations revealed episodic (mildly) relativistic or curved jets roughly in the north-south direction \citep[e.g.][]{Strom1989, Newell1998, Marti2001, Mioduszewski2001, Miller-Jones2004}. Moreover, there are long-term correlations between radio emission and X-ray states \citep[e.g.][]{Szostek2008, Zdziarski2016}. 

To search for a potential physical connection between the radio jet and the X-ray funnel in Cyg~X-3, it is important to observe the innermost jet in the quiescent radio state. To date, it is still difficult for high-resolution very long baseline interferometry (VLBI) imaging observations to reveal the innermost jet structure mainly because of two limitations. As a source in the Galactic plane with a latitude of $b=+0\fdg7$, Cyg~X-3 suffers strong scatter broadening \citep[e.g.][]{Molnar1988, Wilkinson1994}. The observed angular size $\theta_\mathrm{obs}$ shows a strong dependence on the observing frequency $\nu$ (in GHz), $\theta_\mathrm{obs} = 448\, \nu^{-2.09}$~mas \citep{Mioduszewski2001}. This indicates that the source is intrinsically unresolved, i.e. $\theta_\mathrm{obs}=\theta_\mathrm{scat}$, where $\theta_\mathrm{scat}$ represents the contribution from scatter broadening. Because of the apparently extended source structure, the correlation amplitude decreases significantly on the long baselines in particular at low observing frequencies. Moreover, there are no nearby compact and bright calibrators available to reliably run phase-referencing observations at high frequencies \citep[e.g.][]{Miller-Jones2009, Spencer2022}. 

In this Letter, we present the results from an observational campaign with simultaneous X-ray and radio VLBI observations. We describe the observations and data reduction in Section~\ref{sec2}, present the results, interpret the observed innermost jet structure, and discuss some potential implications in Section~\ref{sec3}. 
 
\section{Simultaneous radio and X-ray observations}
\label{sec2}

\subsection{4-Gbps EVN experiment at 22.25 GHz}
\label{sec2:evn}

We observed Cyg~X-3 with the European VLBI Network (EVN) at 22.25~GHz for 2~h (UT 12--14~h) on 2022 November 10. Because of the limitation of scatter broadening \citep[e.g.][]{Spencer2022}, we only required seven European stations to get proper baseline correlation amplitude. The participating telescopes were Jodrell Bank Mk2 (JB2), Effelsberg (EF), Medicina (MC), Noto (NT), Onsala (ON), Yebes (YS), Metsahovi (MH). The shortest baseline is EF-JB and has a baseline length of 700~km. The EVN observations were performed with the maximum date rate 4096~Mbps (16 32-MHz sub-bands per polarization, dual circular polarization and 2-bit quantization). The data correlation was done by the EVN software correlator SFXC \citep[][]{Keimpema2015} at JIVE (Joint Institute for VLBI ERIC) using the typical correlation parameters for continuum experiments: 0.5-MHz frequency resolution and 1-s integration time. 

The observations were performed in the reverse phase-referencing mode. Cyg~X-3 was used as a calibrator to image its nearby ($10\farcm4$) faint source \target{} \citep[source ID S23,][]{Benaglia2021}. From the existing multi-frequency VLA observations (project code 19A-422), we found that the source has a relatively flat spectrum with a flux density of 14--18~mJy at 1.4--7.5~GHz (c.f. Appendix~\ref{app:cal}). We observed the pair of sources with a cycle time of 180~s (100~s for Cyg~X-3 and 40~s for \target{}). All the telescopes had an observing elevation of $\ge30\degr$. Moreover, the bright source J2007$+$4029 was observed as a fringe finder and a bandpass calibrator for two short scans. 

The National Radio Astronomy Observatory (NRAO) Astronomical Image Processing System \citep[\textsc{aips} version 31DEC21,][]{Greisen2003} software package was used to calibrate the visibility data. We followed the EVN data calibration strategy reported in \cite{Yang2022}. All the related \textsc{aips} tasks were called via the \textsc{parseltongue} interface \citep{Kettenis2006} and integrated in a single script. Because of the limitation of 512-MHz bandwidth filters in the digital backend, the top and bottom subbands had low correlation amplitudes and were flagged out. NT had poor fringes in some subbands. These subbands were also excluded.   

The iterations of deconvolution and self-calibration were performed in \textsc{difmap} \citep{Shepherd1994}. Because of the absence of short baselines and the large source structure, the deconvolution was achieved via fitting the visibility data to a Gaussian model instead of using the normal algorithm \textsc{clean}. We tried both circular and elliptical Gaussian models. Compared with the elliptical Gaussian model, the circular model gave about three times higher residuals in the residual map and about three times lower total flux density. Thus, the circular model was rejected in our data analysis. To get the more robust self-calibration solutions for all the stations, we used relatively longer solution intervals: 40~min for the phase self-calibration and 60~min for the phase and amplitude self-calibration. 

The source \target{} is detected with the data on the very short baselines $\leq50$~Mega-wavelengths (hereafter, M$\lambda$). Since it is not the main interest of the Letter, we report these related results and plots in Appendix~\ref{app:cal}.

\subsection{X-ray observations}
\label{sec2:nicer}
\begin{figure}
\centering
\includegraphics[angle=90, width=\columnwidth]{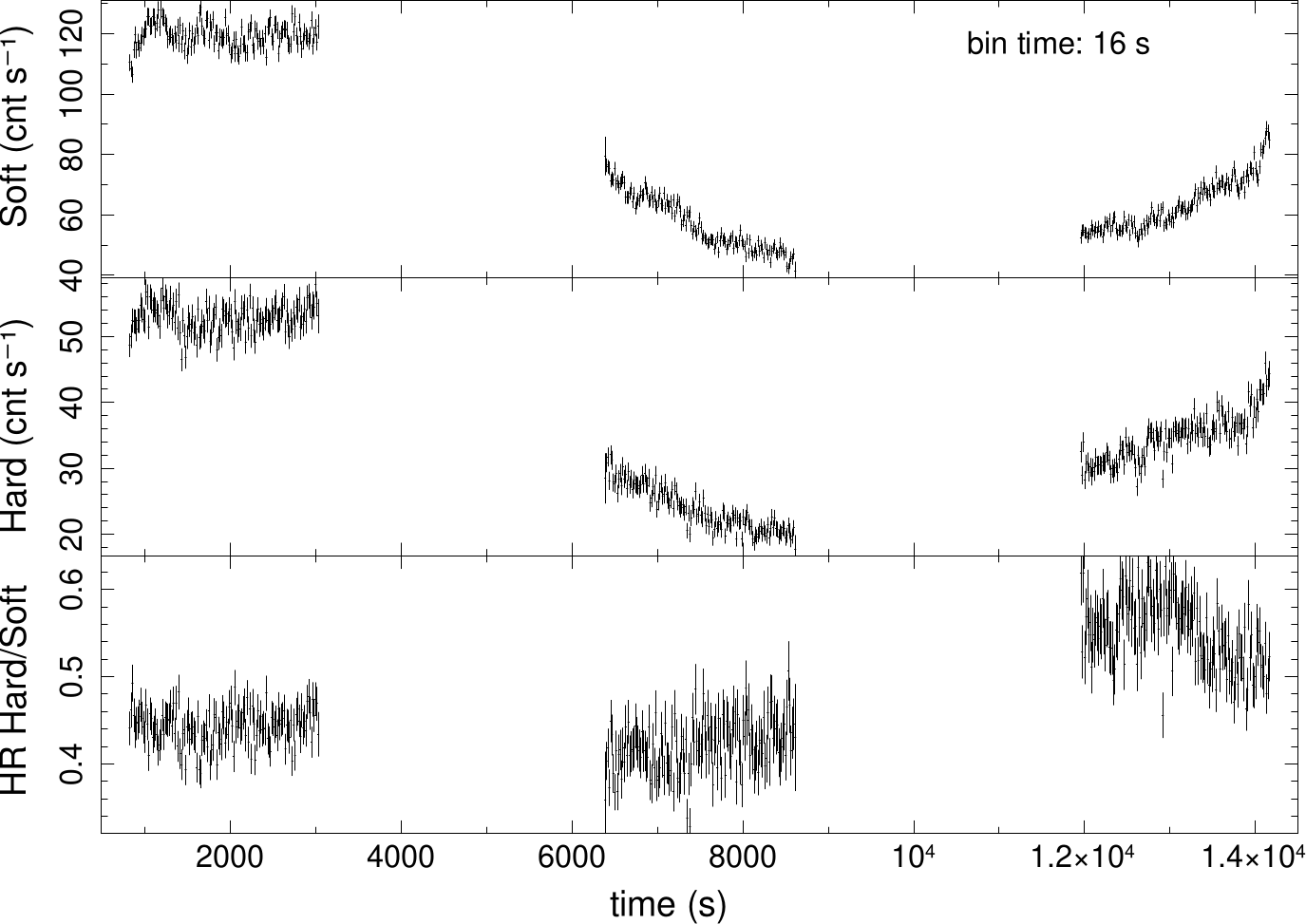}
\caption{
\Ni light curves in soft (1.3--5.0~keV; top panel) and hard (5.0--12.0~keV; middle panel) X-rays. The bottom panel shows the hardness ratio (HR) between the hard and soft bands. Time refers to MJD 59893 11:13:40 UT.}
\label{fig:x-r_lc}
\end{figure}

\begin{figure}
\centering
\includegraphics[clip, trim=0.2cm 1.0cm 2.4cm 1.0cm, angle=0, width=\columnwidth]{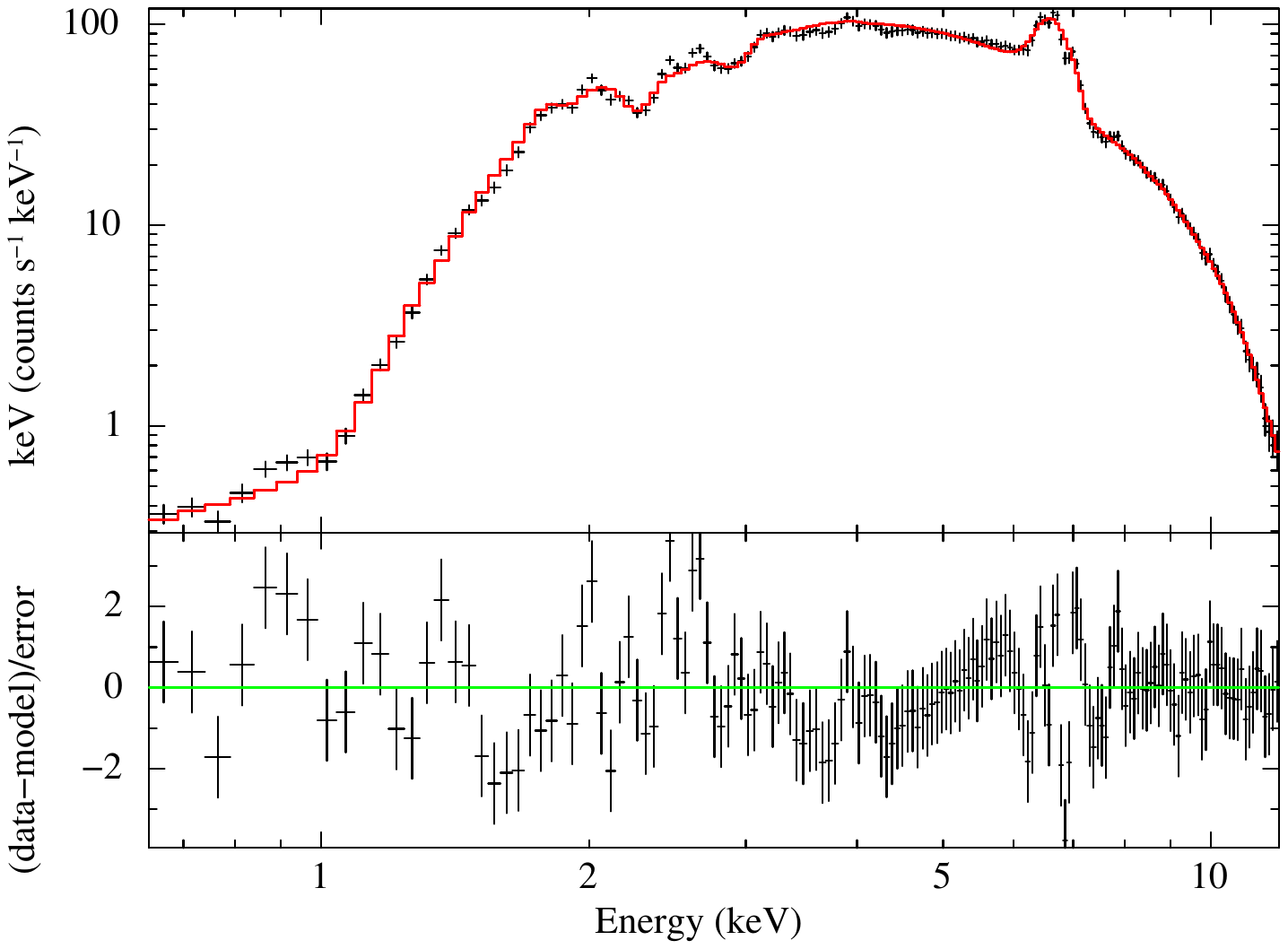}
\caption{\Ni spectrum during the first orbit of the observation, in which the source was steady. In the bottom panel, we show the $\Delta$ residuals of the best-fitting model \textit{TBabs * (reflect*smedge*nthcomp + gauss)} (see text for details). }
\label{fig:x-r_spec}
\end{figure}

We observed Cyg~X-3 with the \Ni X-ray instrument on board the {\it ISS} on 2022 November 10 as part of a director's discretionary time (DDT) request. Observations were carried out in three consecutive 2.2-ks intervals covering the visibility windows immediately before, during, and after the EVN radio observation. \Ni observations started on UT 11:10:57 and ended on UT 14:55:00.

We reduced the observations using \textsc{NICERDAS version 10} in \textsc{HEAsoft version 6.31.1}, together with the latest version of calibration files available (\textsc{CALDB xti20221001}). We reprocessed the observations using \textit{nicerl2} task, discarding detectors 14 and 34. We produced light curves with the task \textit{nicerl3-lc} in two energy ranges: 1.3--5.0~keV (``soft'') and 5.0--12.0~keV (``hard''). These energy ranges were chosen to match those used in \cite{Hjalmarsdotter2008}. The light curves with a binning time of 16~s are shown in Fig.~\ref{fig:x-r_lc}. The light curves present a steady segment followed by two variable segments. The latter corresponds to the well-known 4.8~hr orbital modulation of Cyg~X-3.

In addition, we created a spectrum for the first 2.2-ks segment in which the source is very steady. For this, we used the task \textit{nicerl3-spect} that generates the RMF and ARF matrices, with the background model \textit{3c50} from \Ni. Following \cite{Veledina2023}, we fitted the full 0.5--12~keV spectrum in \textsc{XSPEC} \citep{XSPEC} using a \textit{TBabs * (reflect*smedge*nthcomp + gauss)} model including interstellar absorption, a fully reflected Comptonized continuum, with a smeared edge evinced at $\sim8.8$~keV, and a broad Gaussian profile to model the Fe K complex present in the spectrum. We also try and add a \textit{diskbb} component which happened to fail to improve the spectral fit, as usual in the hard state of Cyg~X-3. For our best-fitting model ($\chi^2$=208 for 147 dof), we obtained a high absorption column of $N_{\rm H} = 3.2\pm0.1\times10^{22}$\,cm$^{-2}$, and an absorbed flux in the 3--5~keV energy range of $F_\mathrm{abs} = (4.46\pm0.03)\times 10^{-10}$\flux. In addition, the unabsorbed flux in the 3--5~keV energy range yielded $F_\mathrm{unabs} = (5.21\pm0.04)\times 10^{-10}$\flux, or equivalently, 0.325(3)~keV\,cm$^{-2}$\,s$^{-1}$. The spectrum and our best-fitting model residuals are shown in Fig.~\ref{fig:x-r_spec}. Despite the known complexity of the Cyg~X-3 spectrum, whose detailed modelling is far from the scope of this work, we find that the spectral state and X-ray flux in this observation is consistent with that of the ``Main'' observation in \cite{Veledina2023}, indicating that the spectral state of Cyg~X-3 at the moment of the EVN observations was coincident with that during the \textit{IXPE} observations. This shows that the IXPE polarization measurements and the EVN radio imaging results are comparable.

\section{Cyg X-3: the innermost jet structure in the X-ray hard state }
\label{sec3}

\begin{table}
\centering
\caption{The best-fitting parameters for the elliptical Gaussian model. The systematic errors were also included in the error budgets.  }
\label{tab:model}
\begin{tabular}{ccccc}
\hline
MJD          
           & Flux         &  Major        & Minor           & PA              \\  
 (d)       & (mJy)        &  (mas)        & (mas)           & ($\degr$)       \\
\hline
59893.54   & $79\pm20$ & $3.18\pm0.03$ & $1.45\pm0.06$   & $-3.1\pm0.4$   \\
\hline
\end{tabular}
\end{table}

\begin{figure}
\centering
\includegraphics[width=\columnwidth]{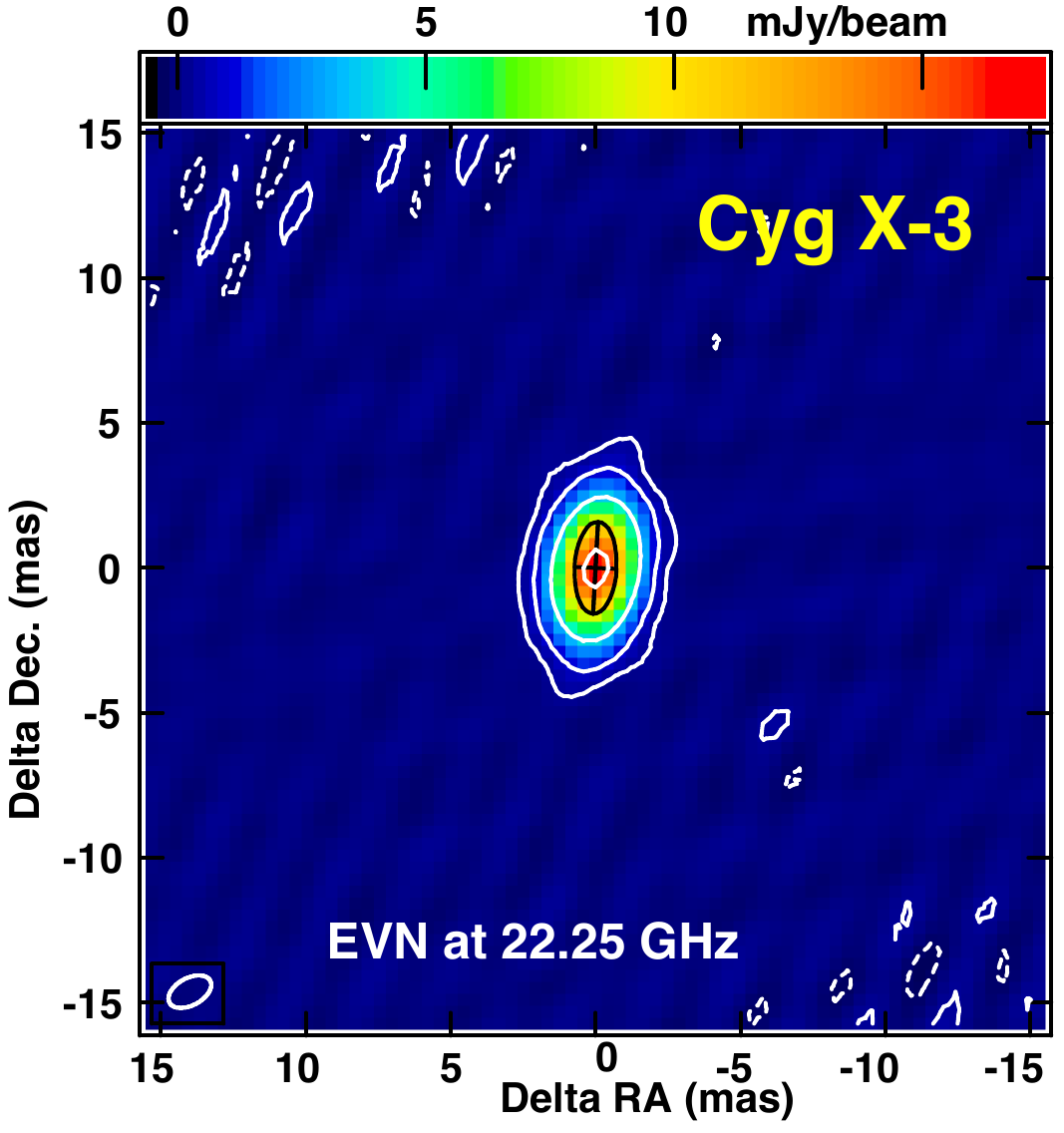}
\caption{
Total intensity map of Cyg~X-3 in the radio quiescent state. The EVN map was observed with seven European telescopes at 22.25~GHz. The black ellipse with a cross shows the best-fitting elliptical Gaussian model reported in Table~\ref{tab:model}. The map used natural weighting and the visibility data of $\le$125~M$\lambda$. The beam full width at half maximum (FWHM) is $1.60\times0.94$ mas at PA~=~$-61\fdg5$ and it is plotted in the bottom-left corner. The white contours plots the levels of ($-$1, 1, 4, 16, 64) $\times$ 3$\sigma_\mathrm{rms}$ ($\sigma_\mathrm{rms}=0.08$~mJy\,beam$^{-1}$). The map peak brightness is 17.3 ~mJy\,beam$^{-1}$. }
\label{fig:evn-map}
\end{figure}

Table~\ref{tab:model} lists the fitting results of the elliptical Gaussian model. Fig.~\ref{fig:evn-map} shows our high-resolution EVN imaging results of Cyg~X-3 in the X-ray hard and radio quiescent state. The source displays an elongated structure at position angle PA=$-3\fdg1\pm0\fdg4$, and has an apparent brightness temperature of $(4.2\pm1.0)\times10^7$~K. Assuming a flat radio spectrum, the VLBI flux density is broadly consistent with the flux densities (average: 76--142~mJy at 1.2--225~GHz) and variability (variance: 12--36~mJy) reported by \citet{Veledina2023} before and after our VLBI observations. During the 2-h EVN observations, the correlation amplitude varied nearly linearly on all the baselines and thus there was no bright flares on timescales $<$1~h. All these radio and the X-ray (c.f. section \ref{sec2:nicer}) properties are fully in agreement with the expectation in the radio quiescent state \citep[e.g.][]{Szostek2008}. Thus, we identify the relatively stable structure as the innermost jet. 
To date, such a VLBI-resolved jet base has also been found in the X-ray binaries: Cyg~X-1 \citep[e.g.][]{Miller-Jones2021} and GRS~1915$+$105 \citep[e.g.][]{Dhawan2000}.   

The plots of the best-fitting model and the visibility data are also shown in Appendix~\ref{app:cygx3}. Because of the absence of short baselines and the relatively poor measurements of antenna gains, the total flux density might have a large systematic error. We used 20\% of the total flux density as the total uncertainty in Table~\ref{tab:model}. There might also exist systematic uncertainties for the observed sizes and PAs because of the poor ($u$, $v$) coverage and potential flux density variability. We also tried to split the 2-h observations into 2--3 short segments and then did the self-calibration and the elliptical Gaussian model fitting. There is no evidence for a change of the source structure. The systematic uncertainties are also likely small or comparable to the formal uncertainties. To give the more reasonable estimates, we enlarged the formal uncertainties by a factor of two and then used them as the total uncertainties in Table~\ref{tab:model}. In the literature, there were a few reports of minor flares resulting from transient ejection activity on timescales up to a few hours \citep[e.g.][]{Molnar1988, Newell1998, Kim2013, Egron2017}. Future EVN observations with longer on-source time and more stations (Sardinia 64-m radio telescope, Robledo 70-m radio telescope, Torun 32-m radio telescope and the enhanced Multi Element Remotely Linked Interferometer Network) would significantly improve the ($u$, $v$) coverage, in particular on short baselines, and thus help to accurately measure its flux density and search for potential small structure changes on short timescales. 

The angular sizes of the major and minor axes in Table~\ref{tab:model} are significantly larger than the prediction ($\sim$0.7~mas) from the scattering broadening model \citep{Mioduszewski2001} and also larger than the minimum size of $\sim$0.7~mas observed by \citet{Molnar1988}, \citet{Kim2013} and \citet{Egron2017} at the same observing frequency during outbursts. Therefore, the jet base of Cyg~X-3 is intrinsically resolved not only in the radial direction, but also in the transverse direction. This has not been seen by the previous VLBI imaging observations mainly because they had relatively lower observing frequencies or low image quality. In the first order, there is a simple relation between the observed size $\theta_\mathrm{obs}$ and the intrinsic size $\theta_\mathrm{int}$, i.e. $\theta_\mathrm{obs} = \sqrt{\theta_\mathrm{int}^2 + \theta_\mathrm{scat}^2}$. If the small contribution of $\theta_\mathrm{scat}$ is excluded and a distance of $7.4\pm1.1$~pc \citep{McCollough2016} is adopted, the innermost jet would have a projected size $(23.5\pm3.5) \times (10.7\pm1.7)$~au. Because of the limited image quality, it is not clear where the compact object is located. Previous high-resolution radio observations during major flares have revealed relativistic jets with two-sided \citep{Marti2001, Miller-Jones2004} or one-sided morphology \citep{Mioduszewski2001} on scales $\ga10$~mas. Depending on flares, the intrinsic speed of short-lived jets could vary from 0.3$c$ \citep{Spencer1986, Shalinski1995, Egron2017} to 0.8$c$ \citep{Mioduszewski2001}. The VLBI structure might represent a two-sided jet as the innermost jets of XRBs in the hard X-ray state are not highly relativistic \citep[e.g.][]{Fender2009}. For example, the XRBs Cyg~X-1 and GRS~1915$+$105 show a relatively symmetric and two-sided jet structure \citep{Dhawan2000, Miller-Jones2021} in the X-ray hard state. The detections of the counter jets near the central objects in these two XRBs indicates that the Doppler beaming effect is weak in the innermost jets.

The innermost jet in Cyg~X-3 had PA~=~$-3\fdg2 \pm 0\fdg4$ on 2022 November 10. This is almost completely orthogonal to the direction (PA: 85$\degr$--95$\degr$) of the linear X-ray polarization observed by \citet{Veledina2023} with the \textit{IXPE} between 2022 October 31 and 2022 December 29. Therefore, the X-ray funnel very likely results from the jet and outflow activity. The coexistence of jets and dusty tori has been frequently observed in active galactic nuclei, e.g. Circinus \citep[][]{Ursini2023} and NGC~1068 \citep[e.g.][]{Gallimore2004}. However, the innermost jets in AGNs generally have a quite compact structure \citep[e.g. a recent review by][]{Blandford2019}. In the case of Cyg~X-3, the wide jet structure might result from its complex environment and orbital motion. There exist strong winds from the companion WR star \citep[e.g.][]{Antokhin2022} that can significantly affect the jet collimation and propagation \citep[e.g.][]{Yoon2016, Bosch-Ramon2016}. This, combined with the orbital motion, leads to a helical jet structure with an increased effective width on scales of 10s to 100s of orbital separations \citep[e.g.][]{Molina2019}. For reference, the orbital separation in Cyg~X-3 is of the order of $\lesssim 2.7\times10^{11}$~cm$\lesssim 0.02$~au \citep{Koljonen2017}, which implies that the measured jet scales corresponds to 100s of orbital separations. The jet could cause some internal shocks and strong interactions with the surrounding high-density envelope. Such internal shocks have been observed in some novae, e.g. V959~Mon \citep{Chomiuk2014}. The jet position angle might also be slightly modulated by the orbital motion. Because of strong scatter broadening, these details on the wind-jet interaction can only be observed by very high-frequency VLBI observations. In the future, the long baseline mode and the 16/32-GHz observing bandwidth of the next-generation Very Large Array (ngVLA\footnote{https://ngvla.nrao.edu/}) would allow us to resolve these inner radio structure at $\sim$90~GHz.

\section*{Acknowledgements}
\label{ack}
The European VLBI Network is a joint facility of independent European, African, Asian, and North American radio astronomy institutes. Scientific results from data presented in this publication are derived from the following EVN project code: RSY09.
This research is based on observations with {\it NICER} (NASA). We thank the {\it NICER PI} and team for promptly approving and scheduling our ToO observations as a DDT.
This research has made use of NASA’s Astrophysics Data System Bibliographic Services.
FG is a CONICET researcher and acknowledges support from PIP 0113 and PIBAA 1275 (CONICET).
WC acknowledges support from the National Natural Science Foundation of China
(NSFC) under grant No. 11903079.
HC acknowledges support from the Hebei Natural Science Foundation of China (Grant No. A2022408002), and the NSFC (Grant No. U2031116).


\section*{Data Availability}
The correlation data underlying this article are available in the EVN Data Archive (\url{http://www.jive.nl/select-experiment}). The calibrated visibility data underlying this article will be shared on reasonable request to the corresponding author. 
The {\it NICER} X-ray data used in this article are available in the HEASARC database (\url{https://heasarc.gsfc.nasa.gov}).




\bibliographystyle{mnras}

\bibliography{Cyg_X-3_MN} 




\appendix
\section{More related figures}
\label{appendix}

\subsection{The radio source near Cyg~X-3}
\label{app:cal}

\begin{figure}
\centering
\includegraphics[width=\columnwidth]{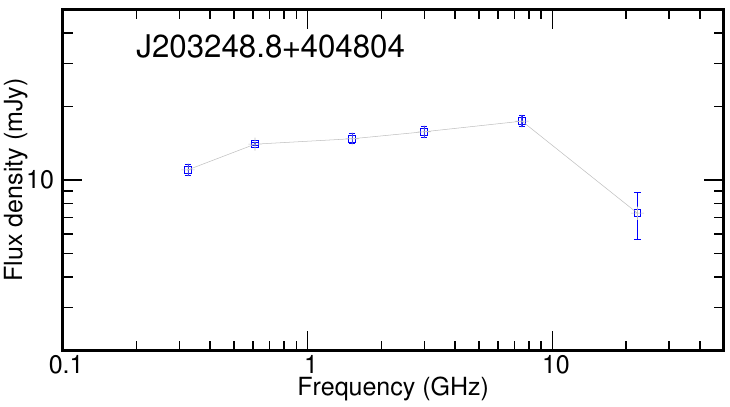}
\caption{
The radio spectrum of the candidate phase-referencing source \target{}.  }
\label{fig:spectrum}
\end{figure}

\begin{figure}
\centering
\includegraphics[width=\columnwidth]{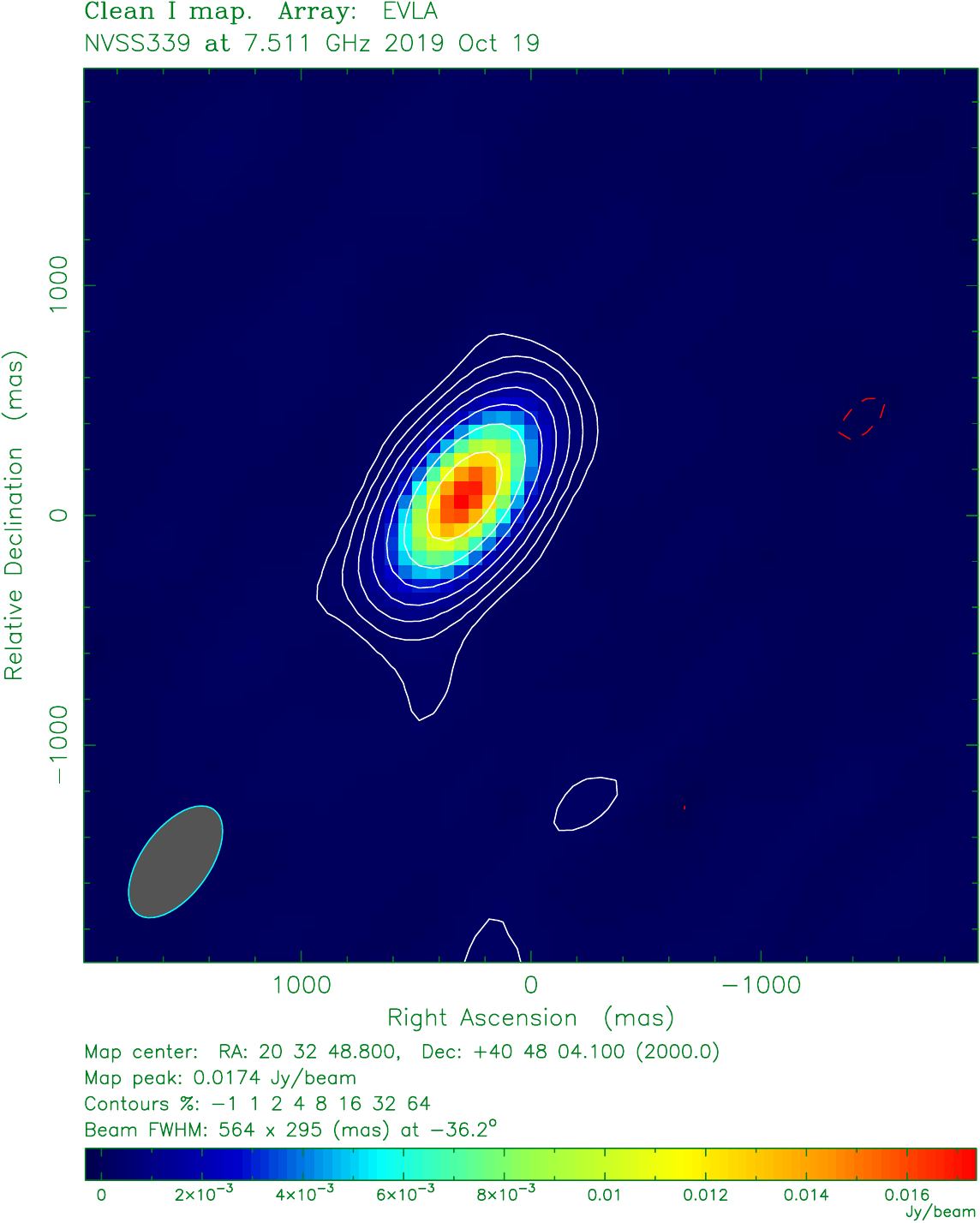}  
\caption{
The arcsec-scale compact structure of the candidate phase-referencing source \target{}. The first contour is at the level $3\sigma_\mathrm{rms}$.  }
\label{fig:vla7-8ghz}
\end{figure}

\begin{figure}
\centering
\includegraphics[width=\columnwidth]{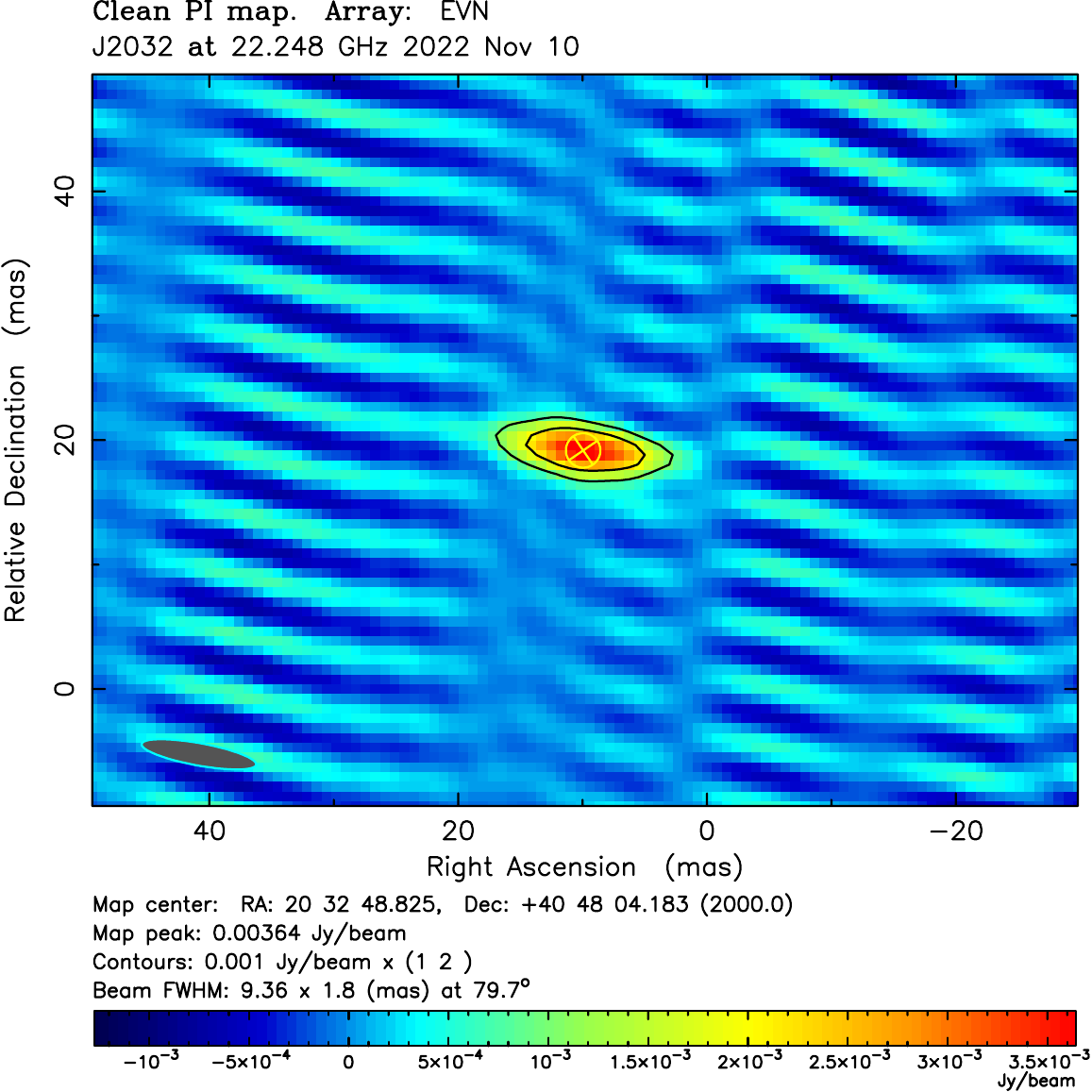}  
\caption{
The EVN detection of the candidate phase-referencing source \target{} on the short baselines $\leq50$~M$\lambda$. The yellow circle represents the FWHM of the circular Gaussian mode. The first contour is at the level $3\sigma_\mathrm{rms}$. }
\label{fig:evn_j2032}
\end{figure}

\begin{figure*}
\centering
\includegraphics[width=\linewidth]{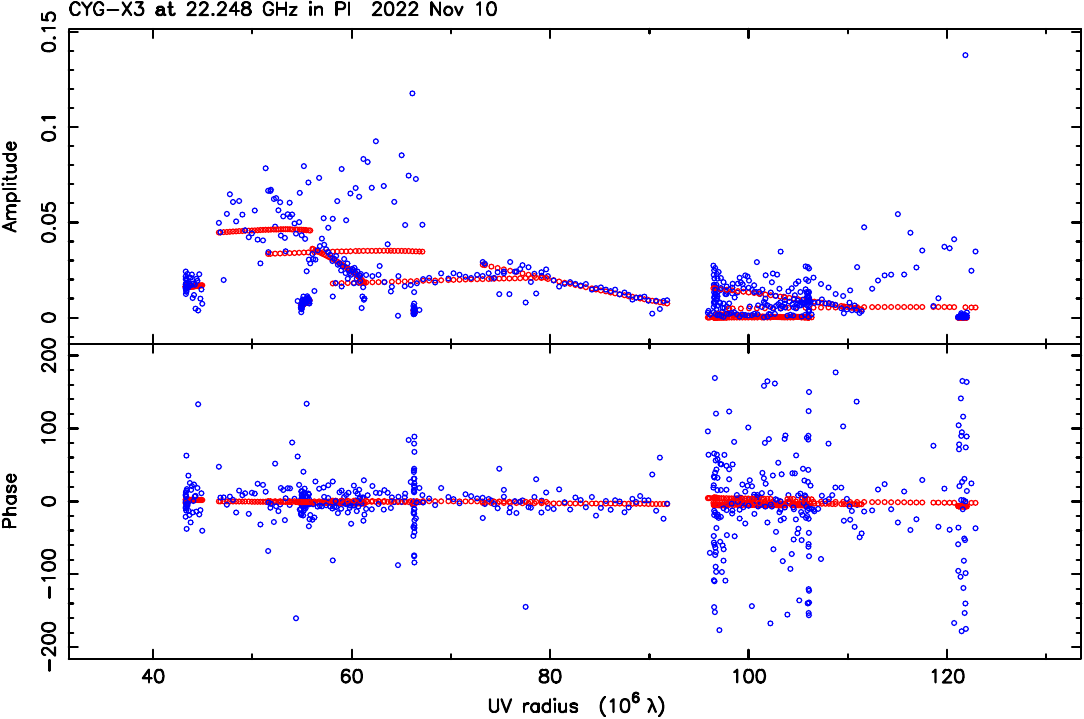}
\caption{
Plots of the baseline correlation amplitude in Jy and phase in degree against ($u$, $v$) radius in M$\lambda$. The blue and red points represent the calibrated visibility data and the elliptical Gaussian model predictions respectively. The data were averaged with all the available subbands and an integration time of 3 min. }
\label{fig:evn-data}
\end{figure*}

The source \target{} is located $10\farcm4$ away from Cyg~X-3.  Because of its closeness, it might be used as a reference point to do VLBI differential astrometry with a precision of $\lesssim$0.01~mas \citep[e.g. a review paper by][]{Rioja2020}. Currently, there are no optical nor infrared counterparts reported in literature. To probe its nature via a broad-band radio spectrum, we analyzed the existing Jansky VLA data (project code 19A--422, PI:  Gregg Hallinan) in the NRAO data archive.  

\target{} was observed by the Jansky VLA at 1--2~GHz, 2--4~GHz, and 4--8~GHz in A configuration on 2019 October 19. During the multi-band observations, \target{} was observed for 2~min per band. The calibrator 3C~48 (B0137$+$331) was observed as the primary flux density calibrator \citep{Perley2017}. The phase-referencing calibrator was J2007$+$4029. The data reduction was performed using the Common Astronomy Software Applications package \citep[\textsc{casa},][]{McMullin2007}. 


Fig.~\ref{fig:spectrum} shows its radio spectrum between 0.3 and 22~GHz. We also added the flux densities at 0.32 and 0.61 GHz reported by \citet{Benaglia2021} in the plot. Fig.~\ref{fig:vla7-8ghz} displays the VLA map at 7--8~GHz. The source is quite compact with a size of $\lesssim$12~mas. The VLA flux density at 2--4 GHz is 15.7$\pm$0.8~mJy, consistent with the measurements ($\sim$14~mJy) from the VLA Sky Survey \citep{Lacy2020, Gordon2021} at the same frequency. Therefore, the source is stable between 2019 and 2022. 

Fig.~\ref{fig:evn_j2032} shows its EVN imaging results. With the very short baselines $\leq50$~M$\lambda$, the natural grid weighting and the unusual beam pattern, the source \target{} is marginally seen at a position close ($<50$~mas) to the image origin in the dirty map. Because of some strong side lobes resulting from poor ($u$, $v$) coverage, it is hard to unambiguously locate the source. The circular Gaussian model fitting shows that the source has a total flux density of $7.3\pm1.6$~mJy, an apparent size of $\sim$2.8~mas and an apparent brightness temperature of $\sim2\times10^6$~K. The VLBI flux density at 22.25 GHz might be underestimated to some degree because of the absence of short baselines. 

In view of the radio spectrum and the high brightness temperature, the source is very likely an extragalactic jet with a partially absorbed radio core. Because it is close to the Galactic plane and has a latitude, $b=+0\fdg55$, it will suffer strong scatter broadening. The apparent large size might have a large contribution from the scatter broadening. 


\subsection{Cyg~X-3: the elliptical Gaussian model fitting}
\label{app:cygx3}

Fig.~\ref{fig:evn-data} plots the calibrated visibility data and the best-fitting elliptical Gaussian model in \textsc{difmap}. The data on the baselines to the most sensitive station EF had significantly smaller scatter. The model tracks the variation of the baseline correlation amplitude. Because the data on the baselines to JB2, NT and MH have limited baseline sensitivity, they have a large scatter and most of them lie above the model. This is as expected because the correlation amplitude follows a Rice distribution instead of a Gaussian distribution. These low-sensitivity data points have very low weights in the self-calibration and the model fitting. We also tried to exclude the three stations in the deconvolution and the results were not changed significantly. It is difficult to accurately measure total flux densities of the resolved source on short timescales due to the lack of $<42$~M$\lambda$ baselines.

\bsp	
\label{lastpage}
\end{document}